\def\etal{{\it et al.~\/}}

\def\ie{{\it i.e.~\/}}

\def\ltsima{$\; \buildrel < \over \sim \;$}
\def\simlt{\lower.5ex\hbox{\ltsima}}
\def\gtsima{$\; \buildrel > \over \sim \;$}
\def\simgt{\lower.5ex\hbox{\gtsima}}
 
\documentstyle[aasms]{article}
\tighten
\singlespace

\lefthead{Ricotti, Ferrara \& Miniati}
\righthead{Energy dissipation in interstellar cloud collisions}

\begin{document}

\title{Energy dissipation in interstellar \\
cloud collisions}

\author{Massimo Ricotti$^1$, Andrea Ferrara$^2$, and Francesco Miniati$^3$}
\affil{
$^1$Dipartimento di Astronomia, Universit\`a di Firenze, \\
50125 Firenze, Italy 
\\ E--mail: ricotti@arcetri.astro.it\\
$^2$Osservatorio Astrofisico di Arcetri \\ 50125 Firenze, Italy 
\\ E--mail: ferrara@arcetri.astro.it \\
$^3$University of Minnesota, School of Physics \& Astronomy\\
Minneapolis, MN 55455, USA
\\ E--mail: min@astro.spa.umn.edu}
%.................................................................

\begin{abstract}
{We present a study of the kinetic energy dissipation in
interstellar cloud collisions. The main aim is to understand the
dependence of the elasticity (defined as the ratio of the final to 
the initial kinetic energy of the clouds) on the velocity and mass ratio of the
colliding clouds, magnetic field strength, and gas metallicity for head-on
collisions.  The problem has been studied both analytically and 
via numerical simulations. We have derived handy analytical relationships that 
well approximate the analogous numerical results.
The main findings of this work are: $(i)$ the kinetic energy dissipation in 
cloud collisions is minimum  (\ie the collision elasticity is maximum) 
for a cloud relative velocity $v_r \simeq 30$ km s$^{-1}$;
$(ii)$ the above minimum value is proportional $Z L_c^2$, where $Z$ is the 
metallicity and $L_c$ is the cloud size: the larger is $ZL_c^2$ the  
more dissipative (\ie  inelastic) the collision will be; $(iii)$ in general,
we find that the energy dissipation decreases when the magnetic field 
strength, and mass ratio of the clouds are increased and the metallicity is 
decreased, respectively.  We briefly discuss the relevance of this study to the 
global structure of the interstellar medium and to galaxy formation 
and evolution.} 

\end{abstract}


\section{Introduction}

The complex morphology and the variety of scales
observed witness to the turbulent nature of the ISM. 
Nevertheless, the vast majority of the dynamic models of the ISM
(but see Norman \& Ferrara 1996, Vazquez-Semadeni \etal 1995)  
describe this medium in terms of 
dense, neutral clouds (CNM) moving into a 
smooth, diffuse intercloud medium (WNM) by which are pressure confined.
If we assume that the cloud velocities are randomly distributed,
a cloud undergoes a supersonic collision 
approximately every $10^6-10^7$ years. Therefore, in the Galaxy,
assuming a volume of the gaseous disk to be $\sim 78$ kpc$^3$, about $0.3 -
3$ cloud collisions occur every 100 years (we have assumed that a 
line of sight through the Galactic disk intersects $\sim 7$ clouds per kpc
[Spitzer, 1978]).

Cloud collisions play an important r\^ole in the evolution
of a galaxy.
For example, the large gas compression following an inelastic
collision might be expected to enhance the star-formation rate (Gilden,
1984).
Inelastic cloud collisions are important for the
global energy budget of the ISM: as the energy input from SNe and HII
regions accelerate the interstellar gas, the cloud kinetic energy is
dissipated by cloud collisions (Spitzer, 1978).
Also, since the
vertical distribution of the HI in the Galaxy 
depends on the turbulent motions of the clouds (McKee 1990, Ferrara 1993), the structure
of the gas is strongly regulated by the dissipation of bulk motions.

In addition, the buildup of the mass spectra of an ensemble of clouds is 
partially affected
by cloud collisions (Oort 1954, Field \& Saslaw 1965, Field \& Hutchings 1968,  
Penston {\it et al.} 1969, Cowie 1980, Hausman 1982, Pumphrey \&
Scalo 1983, Struck-Marcell \& Scalo 1984, Fleck 1996).
The basic idea is that small clouds, due to inelastic 
collisions, coalesce to form larger ones which eventually collapse to form stars.
Klein, McKee \& Woods (1995) have recently pointed out
that the outcome of an inelastic collision may be shattering rather than
coalescence due to the fast growth  of a nonlinear instability, the so-called
bending mode instability.  This result has a significant impact on the cloud 
buildup models. Moreover, the morphology of the clouds after the collision event is
strongly dependent on both the elasticity (defined as the ratio of the final to
the initial kinetic energy of the clouds) and the presence of a magnetic
field. 

The rationale for the assumption of inelastic collisions and  
subsequent coalescence of the colliding clouds commonly adopted by the buildup
models can be summarized as follows.
Interstellar clouds motions are highly supersonic; therefore, during
collisional events, two shock waves arise and propagate from the contact
discontinuity (CD) heating the gas; this hot gas, being
optically thin, tends to cool quickly due to radiative losses.
For a standard diffuse cloud,
the typical cooling time of the shocked gas is 
much less that the characteristic timescale of the collision. This justifies
the inelastic approximation.
Nevertheless, this assumption might be rather crude for small clouds and/or 
primordial galaxies where the gas metallicity is low and therefore the cooling
time is longer. 

In this work we will investigate the physics of interstellar cloud
collisions,
with particular emphasis devoted to the dependence of the kinetic energy dissipation
on the parameters of the problem, such as the cloud relative
velocity, gas metallicity, cloud mass ratio, and magnetic field
strength.
The final goal is to identify the regions of the parameter
space in which the collisions are elastic (adiabatic), inelastic 
(isothermal) or radiative. 

The plan of the paper is as follows. In \S~II we discuss the main
heating/cooling processes of the ISM and calculate its thermal balance. 
In \S~III, we will present a first analytical approach 
that will allow us to find approximate relationships between the
elasticity and the cloud parameters
(relative velocity, sizes, gas metallicity, dust-to-gas
ratio).
These analytical results should be useful for 
future works concerned with a statistical approach to
the dynamics of the galactic ISM, 
the mass spectrum of diffuse clouds, and star formation processes.
In \S~IV, the results of a series of numerical simulations of head-on collisions 
are considered; 
in \S~V we will discuss the results and the dependence of the
elasticity on the various parameters of the collision comparing the analytical and
numerical approaches. A brief summary closes the paper.

\section{Thermal balance of ISM}

In this Section we calculate the thermal equilibrium of the diffuse
interstellar medium that is illuminated by the local interstellar
far-ultraviolet and X-ray radiation field and permeated by the
cosmic-ray flow.
We follow closely the work of Wolfire \etal  (1995),
which have incorporated the photoelectric heating from small grains and
PAHs (with a distribution of sizes in the range  3--100 \AA).
Considering a  realistic cooling/heating function is important for two reasons:
(i) we want to study the energy dissipation in collisions occurring in a 
two-phase medium; (ii) the elasticity of the collision is closely connected 
to the emission processes. In addition we would like to explore the effects of 
gas metallicity, $Z$, and dust-to-gas ratio, $D/G$ (normalized to the local
value), on the equilibrium states of the ISM. 
A summary of the dominant cooling processes included in our function is 
given in Table 1. The Table is intended to complement the analogous one
in Wolfire \etal (1995).

In order to obtain an analytical expression for the fractional ionization of 
H and He, and the electron density $n_e$, we have introduced the approximations
described below.  The motivation for this choice is to decrease the computational 
time of the numerical simulations which make use of the cooling function. 
We assume the ionization equilibrium and we neglect: 
recombination on dust grains,
He dielectronic recombination,
ionization due to hydrogen collisions,
secondary ionization by cosmic-rays;
%$\bullet$ in the parametric expression of the ionization due to X-ray we fix 
%$n_e=10^{-2}$ cm$^{-3}$ (for this expression see Wolfire \etal (1995)
%Appendix A). 
in addition, we assume $x_{H^+}=n_{H^+}/n =
x_{He^+}=n_{He^+}/n(He)$, where $n=n_{H}+n_{H^+}$,
$n(He)=n_{He}+n_{He^+}$ and $n(He)/n=0.1$.
With these approximations the equation of the ionization equilibrium can
be solved analytically to obtain an expression for $n_e(n,T)$.

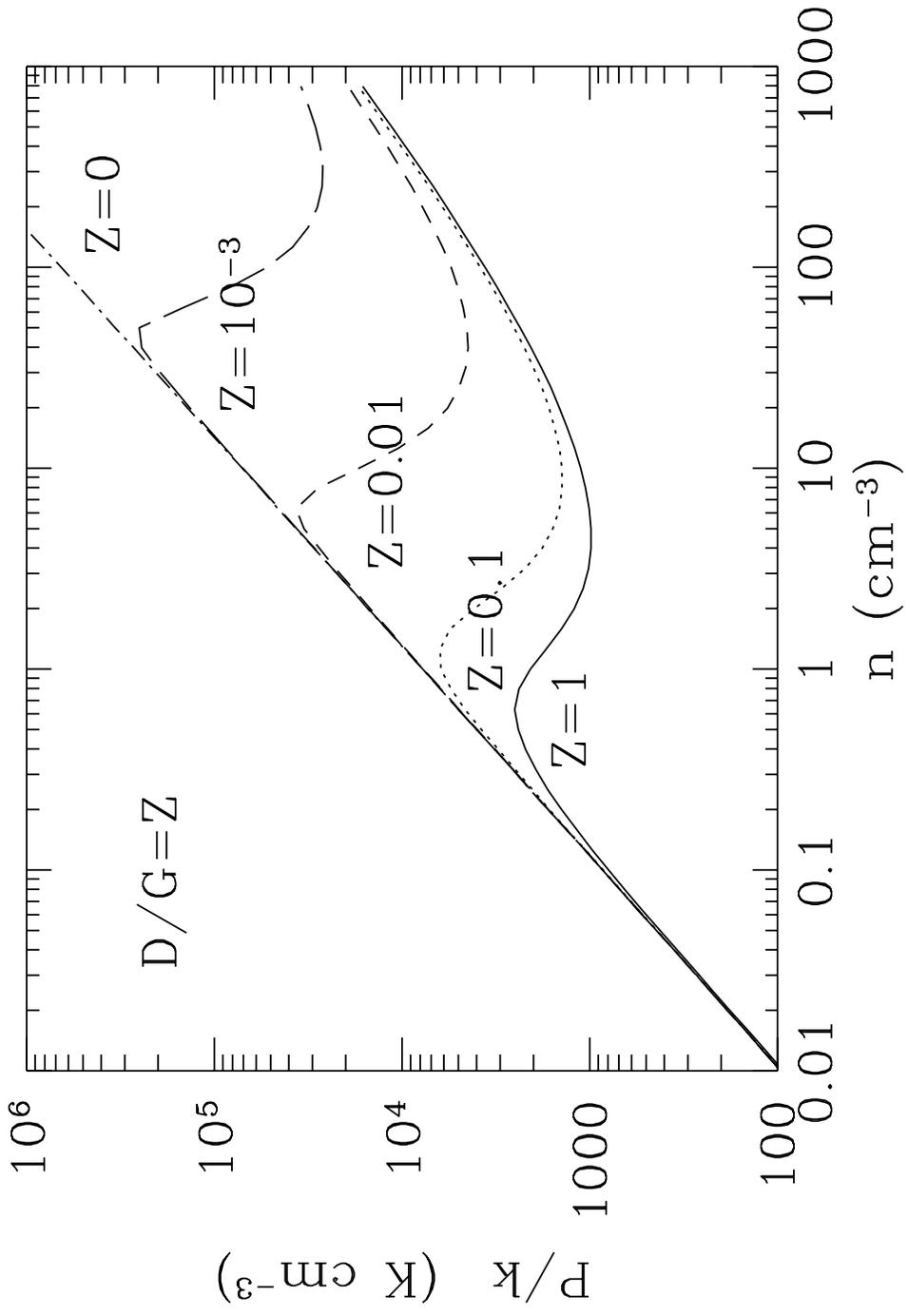
\begin{figure}
\caption{\label{fig1} Thermal pressure $P/k$ vs. hydrogen density
$n$; the curves refer to different values of the 
dust-to-gas ratio, $D/G$, and metallicity $Z$, with $D/G
= Z$. The gas is thermally stable for $(d \log P/d \log n)>0$
(\ie positive slope of the curves).
Unless $Z=0$, a stable two-phase medium is supported.} 
\end{figure}

In Figure~\ref{fig1} we show the phase diagram (pressure versus density)
of the gas for different values of the metallicity
and of the dust-to-gas ratio.
When $Z$ (which we will assume to be equal to $D/G$ throughout the paper) is decreased 
the pressure range in which a multi-phase medium can
exist becomes wider; in addition the mean equilibrium pressure and density of 
the clouds increase.
The photoelectric heating of PAHs and small grains dominate on the
cosmic-ray and X-ray heating both in the intercloud medium (WNM) and in the
clouds (CNM).
The WNM is cooled mainly by the emission of Ly$\alpha$, CII (158
$\mu$m) and
OI (63 $\mu$m). The CNM is cooled mainly by CII (158
$\mu$m). 

\section{Analytical approach}

Before we describe the results of the numerical simulations, we present here
a simple analytical model for the collision. As already stated in the
Introduction, the motivation is twofold: on the one hand we can easily isolate 
the most important physical processes; on the other hand simple approximated
relationships can be found that can be used to explore a wide region of 
the parameter space.

As a first approach to a parametric study of the energy dissipation in
interstellar cloud collisions we focus on a 1D model, which
allows us to consider centered collisions only; the study
of off-center collisions is devoted to a future paper.
We will initially assume that the two clouds are identical  and we will
release this assumption in Sec. 3.4.  In this simple scenario, clouds are 
modeled as cylinders of base area $S_c$ and initial length $L_c$.
The time origin $t=0$ is defined as the moment at which the bases of the 
cylinders coincide in a plane, \ie the contact discontinuity.
We use a reference frame (LSR) with the $x$-axis along 
the symmetry axis of the cylinders and the  $y-z$ plane coinciding with the
 CD.
In the LSR,  $v_c$ is the cloud velocity, $v_f$ is the shock velocity
(see below), and $v_{CD}$ the velocity of the CD 
(for the collision of two identical clouds, $v_{CD}=0$, and for the 
symmetry of the problem we can consider the evolution of one cloud only);
the clouds have initial density $n_0$, and a mass
$M_0=S_c L_c \mu m_H n_0$,
where $\mu$ is the mean molecular weight of the gas and $m_H$ is the
proton mass.

Three main evolutionary stages of the collision can be identified:
(Stone, 1970a,b):
(1) a {\it compression} phase: two shocks propagate from the CD into the clouds 
converting the kinetic energy of the unperturbed
gas into internal energy of the shocked gas. At the end of this phase the
clouds are crushed into a disk whose thickness depends on the entity of
radiative losses. 
(2) an {\it expansion} phase: when the shock reaches the rear side the cloud expands
because of the decrease of the ram pressure of the incoming flow. 
The shock will propagate into the intercloud medium and a
rarefaction wave will propagate backwards in the cloud.
(3) a {\it collapse} phase: the rarefaction wave brings the pressure of the
cloud below the ambient pressure: when the rarefaction wave is reflected
by the cloud leading edge it becomes a compression wave thus halting the
expansion. In this phase the cloud rear edge becomes Rayleigh-Taylor unstable.

%For a supersonic collision we find
%from a dimensional analysis (Gilden, 1984):
%\begin{equation}
%t_{ra} \ll t_{coll} \ll t_{ex}
%\end{equation}
%in the isothermal limit, and 
%\begin{equation}
%t_{ra} \sim t_{coll} \sim t_{ex}
%\end{equation}
%in the adiabatic limit.
%Where $t_{ra}$, $t_{coll}$ and $t_{jet}$ are the typical times of the expansion
%phase, of the compression phase and lateral expansion, respectively.
%In the isothermal case the collision is effectively almost unidimensional.
%************

Depending on the values of $N_{cool}=n_0 v_c t_{cool}$, the column density of the
post-shock radiative region (where $t_{cool}$ is the typical cooling time of the
shocked gas), and $N_c=n_0L_c$, the total cloud column density, the collision will 
be defined to be:
(a) adiabatic: $N_{cool} \gg n_0 L_c$ ($t_{cool} \gg t_{coll}$);
(b) isothermal: $N_{cool} \ll n_0 L_c$ ($t_{cool} \ll t_{coll}$);
(c) radiative: $N_{cool} \sim n_0 L_c$ ($t_{cool} \sim t_{coll}$);
where $t_{coll}=L_c/v_c$ is the typical collision time.
In the following we will briefly discuss the limiting cases (a) and (b)
and present in more detail the radiative case.

\subsection{Adiabatic collisions}

The shock wave generated by the collision event will propagate from the CD through 
the cloud with a velocity $v_f$ in the LSR and $v_s=v_c+v_f$ with
respect to the 
unperturbed gas; the shocked gas is at rest in the LSR and
moves with velocity $v_f$
with respect to the shock front.
For a strong shock in a perfect gas with $\gamma=5/3$,  the density
$\rho_1$, the pressure, $p_1$, the temperature, $T_1$ and the sound speed,
$C_1$, of the postshock gas are: 
\begin{equation}
\label{an_02}
{\rho_1 \over \rho_0}={v_s \over v_f}\simeq 4,~~~~~~
{P_1 \over P_0}\simeq {5 \over 4} {\cal M}^2,~~~~~~
{T_1 \over T_0} \simeq {5 \over 16} {\cal M}^2,
C_1 \simeq {\sqrt 5 \over 4} v_s.
\end{equation}
At the end of the compression phase,  $t \sim t_{coll}$,  the gas density is 
enhanced by a factor 4, the shock 
velocity is $v_s\simeq {(4 / 3)} v_c$, and the cloud
length is $L\simeq L_c / 4$.
By definition, for an adiabatic collision the radiated energy, $E_r$,  is zero:
\begin{equation}
{E_r \over E_k^i}=0,
\end{equation}
where $E_k^i=Mv_c^2$ is the initial kinetic energy of the clouds.
The elasticity of the collision is:
\begin{equation}
\epsilon={E_k^f \over E_k^i} \simeq {9 \over 4} {C_1^2 \over v_c^2}
\simeq 1.
\end{equation}
where $E_k^f=Mv_f^2\simeq (9/4)M C_1^2$ is the kinetic energy of the 
cloud in the expansion phase.

\subsection{Isothermal collisions}

In an analogous manner, for isothermal collisions we find:
\begin{equation}
\label{an_03}
{P_1 \over P_0}\simeq {\rho_1 \over \rho_0}={v_s \over v_f}\simeq
{\cal M}^2,~~~~~~
{T_1 \over T_0} ={C_1 \over C_0} \simeq 1.
\end{equation}
As a consequence, when the shock reaches the rear side of the cloud,
the cloud size is reduced to $L\simeq L_c / {\cal M}^2$ and the
shock velocity is $v_s\simeq[{{\cal M}^2/({\cal M}^2-1)}]v_c \sim v_c$.
In this case the initial kinetic energy will be almost completely 
dissipated and $\epsilon \simeq {\cal M}^{-2} \sim 0$.

\subsection{Radiative collisions between identical clouds}

In the intermediate cases in which $N_{cool} \sim n_0L_c$
(radiative collisions), it can be shown  that the elasticity of the
collision depends on a single parameter $\eta \propto
N_{cool}/(n_0L_c)$; we will refer to $\eta$ as the ``elasticity parameter''.
After the shock has reached the rear side of the cloud, the cloud begins an
expansion phase that we will assume to be adiabatic. This assumption is
reasonable since the typical expansion timescale is much shorter than 
$t_{cool}$.

At this time there are two physically different
regions in the cloud interior: (i) an adiabatic layer behind the shock 
front; we will assume
that the column density of this region is proportional to the column
density $N_{cool}$ of the radiative region of the shock.
(ii) an isothermal region in which the temperature of the gas is
approximately in thermal equilibrium.

As we have seen in the previous Section, 
the major contribution to the radiated energy comes from the isothermal region,
whereas the kinetic energy is essentially stored into the adiabatic region.
If we define the ``elasticity parameter'' of the collision , $\eta$, as
\begin{equation}
\eta = \alpha \cdot {N_{cool} \over n_0 L_c},
\label{parametro_el}
\end{equation}
where $\alpha$ is defined by $N_{ad}=\alpha \cdot N_{cool}$ 
(\ie we assume that the column density of 
the adiabatic region is proportional to the cooling column density)
we obtain the following relationship for the radiated energy and collision
elasticity $\epsilon$:
\begin{eqnarray}
1-\epsilon={E_r \over E_k^i} &\simeq& ( 1-\eta ), \\
\epsilon={E_k^f \over E_k^i} &\simeq& \eta;
\end{eqnarray}
obviously, $\eta \ll 1$ corresponds to isothermal collisions; $\eta=1$ denotes
adiabatic collisions. 

From a dimensional analysis, we find:
\begin{equation}
t_{cool}={3 \over 2} (1.1+x_e^{eq}){n k\delta T \over
{\cal L}(x_e^{eq},n^{eq},T)}.
\label{eq_tcool}
\end{equation}
where $\delta T=T-T^{eq}$.
Substitution of  eq.  (\ref{eq_tcool}) into eq.  (\ref{parametro_el})
yields, 

\begin{equation}
\eta \propto {v_s n_0 \delta T \over L_c \vert {\cal L}\vert},
\end{equation}

Behind the shock front it is $\vert {\cal
L}\vert=\vert n_0^2\Lambda(T)-n_0\Gamma(T)\vert \simeq n_0^2\Lambda(T)$,
because the radiative losses dominate over the heating and $\delta T
\propto v_s^2$ for the adiabatic Rankine-Hugoniot jump conditions. It follows
that
\begin{equation}
\eta \propto {v_s^3 \over n_0 L_c \Lambda(v_s,Z)}.
\label{eq_eta1}
\end{equation}
In Figs.~\ref{fig4} and \ref{fig5} we show (solid lines) the dependence of
$(1-\eta)=E_r/E_k^i$ on  the relative velocity, $v_r$, of the clouds for
different values of $L_c$, and $Z$ as given by eq. \ref{eq_eta1}.  
In Sec. 5 we will show how the above analytical solutions approximate quite
well the analogous results obtained from the numerical simulations.
We have assumed that $v_s \sim v_c=v_r/2$ (this hypothesis holds exactly for a
completely inelastic collision, see Sec. 3.2).
The number density $n_0$ and temperature $T_0$ of the unperturbed
clouds are computed imposing pressure equilibrium with the
intercloud medium; if $D/G=Z=1$ we assume $P/k=2300$ cm$ ^{-3} $ K for
the equilibrium pressure.
If $Z \not= 1$, the equilibrium pressure is computed fixing the CNM
temperature to the value found with $D/G=Z=1$, $T_0 \simeq 48$~K;
thus, if the metallicity decreases, clouds become more
dense and the equilibrium pressure increases.
The motivation for this choice is that the equilibrium temperature of the
clouds is poorly sensitive to the value of the metallicity and pressure 
of the ISM.

\subsection{Radiative collisions between different clouds}

We now consider colliding clouds with different velocities and
sizes; the indexes $1$ and $2$ refer to parameters of the 
left and right cloud, respectively.  We define the nondimensional parameters:
\begin{equation}
\kappa={L_{c1} \over L_{c2}}={M_1 \over M_2},~~~~
\beta=\left \vert {v_1 \over v_2} \right \vert,
\end{equation}
Since we have assumed the same number density $n_0$ for the two clouds the size
ratio is the same as the mass ratio; 
$v_{CD}$, is computed imposing the
same ram pressure on  the two sides of the discontinuity.
Until the shock wave is inside the smaller cloud we have:
\begin{equation}
\left \vert v_{CD}\right \vert =\left \vert {v_2 \over 2}
(\beta -1)\right \vert.
\end{equation}
In the frame of reference comoving with the CD 
(DSR),  the cloud velocity will be:

\begin{equation}
\vert v_1^\prime\vert  =\vert  v_2^\prime\vert  ={v_r \over 2}=\vert
v_2\vert (\beta+1),
\end{equation}
where $v_r$ is the relative velocity of the clouds.
This phase of the collision (in the DSR) is the same as for
two identical clouds.
The kinetic energy in DSR, $E_k^\prime$, is related to the kinetic
energy in the LSR, $E_k^i$, by the relationship:
\begin{equation}
E_k^\prime = f(\kappa, \beta) \cdot E_k^i,
\end{equation}
where
\begin{equation}
f(\kappa , \beta)={ (\beta+1)^2 (\kappa+1) \over 4 (1+\kappa \beta^2)};
\end{equation}
obviously if $\kappa=\beta=1$ then $f(1,1)=1$.
Since $E_{r}$ is the same in both reference frames, then the
elasticity in the LSR is just the product of the elasticity in the DSR and  
$f(\kappa ,\beta)$.

As the shock wave exits from the smaller cloud, the discontinuity starts 
to move (in the DSR) under the effect of the ram pressure of the bigger cloud.
The dynamics becomes rather complex since $v_{CD}$ depends on the
ram pressure ($\propto v_s^2$) and $v_s$ depends on 
$v_{CD}$ as 
$v_s(t)=v_s(0)-v_{CD}(t)$ (in the DSR).
During this phase the shock velocity is not constant and the radiated energy
must be derived by integration:
\begin{eqnarray}
{E_r \over E_k^i} \propto {1 \over S_cN_cv_c^2}\int_0^{t_{coll}} dt~n_0^2
\Lambda S_c L_{r}&=&{1 \over N_cv_c^2}
\int_0^{N_{c}-\alpha N_{cool}} dN~v_s^2=\nonumber \\
\left({\overline v_s^2 \over v_c^2}\right)
{N_c-\alpha N_{cool} \over N_c}&=&\xi (1-\eta),
\label{eq_nub_div}
\end{eqnarray}
where
\begin{equation}
\overline v_s^2={1 \over N_c-\alpha N_{cool}} \int_0^{N_{c}-\alpha
N_{cool}} dN~v_s^2
=\xi v_s^2,
\label{eq_int_csi}
\end{equation}
where $L_{r}$ is the length of the post-shock radiative region,
$\alpha N_{cool}$ is the column density of the adiabatic region at the
moment in which the shock exits from the cloud. 
If $v_s(t)=cost.$, then $\xi=1$ and we recover the case of identical
clouds: $E_r/E_k^i=(1-\eta)$.
After some algebra we find (in the DSR):
\begin{equation}
{E_r \over E_k^i}=\left[{2-\xi \over 1+\kappa}(1-\eta_2)+{\kappa \xi
\over 1+\kappa}(1-\eta_1)\right],
\end{equation}
where
\begin{equation}
\eta_1={N_{cool}(v_{s1}) \over N_{c1}},~~~~\eta_2={N_{cool}(v_{s2})
\over N_{c2}},
\end{equation}
$v_{si}$ is the shock velocity exiting from cloud $i$, 
$N_{ci}$ is the column density of cloud $i$.
If we assume that $v_{CD}\sim v_{CM}/2$, where $v_{CM}$ is the velocity
of the center of mass, at the end of the collision, we
have: $v_{s2}\simeq v_c$ and $v_{s1}\simeq v_c(\kappa+3)/[2(\kappa+1)]$.
We derive $\xi$ from an energy conservation argument: 
if all the kinetic energy relative to the center of mass is dissipated we
obtain $\eta_1=\eta_2=0$. In this case $E_r/E_k^i=[2+\xi (\kappa
-1)]/(1+\kappa)$ is equal to $1-(E_{CM}/E_k^i)=4 \kappa/(\kappa +1)^2$,
where $E_{CM}$ is the kinetic energy of the center of mass.
This allows us to conclude that  $\xi=2/(\kappa+1)$; finally 
in the LSR we have:
\begin{equation}
1-\epsilon = {E_r \over E_k^i}=f(\kappa,\beta)\left[{2-\xi \over
1+\kappa}(1-\eta_2)+{\kappa\xi
\over 1+\kappa}(1-\eta_1)\right].
\end{equation}

\subsection{Lateral outflow}

In a realistic situation clouds have a finite size perpendicular to the
collision direction ($x$-axis); thus the gas will be free to expand in
the directions parallel to the shock front, possibly generating a lateral
outflow.
We will compute the amount of mass lost in the lateral outflow analytically.
To this purpose, we integrate the flux of mass from the cloud surface 
on the typical collision time $t_{coll}= L_c /v_c$.
The typical expansion velocity of a gas in the vacuum is proportional to
the sound speed; as a consequence, most of the mass loss comes from the 
hotter post-shock adiabatic region.
During the compression phase the amount of mass ejected in the lateral outflow 
will be:
\begin{equation}
M_{out} = \pi L_y L_{ad} v_{out}t_{coll} 4 \rho_0,
\end{equation}
where $L_y$ is the cloud diameter (clouds are modeled as cylinders), 
$L_{ad}$ if the length of the adiabatic region, $v_{out}$ is the outflow
velocity, and $4 \rho_0$ is the post-shock 
gas density.
After some algebra, and recalling   that $v_{out} \simeq v_s$ and
$4 \rho_0 L_{ad} \equiv \alpha N_{cool}$, we find:
\begin{equation}
{M_{out} \over M_c} \simeq \eta_y.
\end{equation}
where $\eta_y=N_{cool}/(n_0L_y)$.
The kinetic energy, $E_{k,out} \propto M_{out} v_{out}^2$, lost in the lateral
outflow is:
\begin{equation}
\label{an_eex}
{E_{k,out} \over E_k^i} \simeq \eta_y.
\end{equation}
The kinetic energy lost in the lateral outflow depends both 
on the elasticity and the geometry of the collision. Since $\eta_y= \eta
(L_c/L_y)$,  
(a) for spherical clouds ($L_c/L_y=1$), the kinetic energy lost in the
lateral outflow is proportional to the elasticity parameter (\ie is zero in a
perfectly inelastic collision and 1 in a perfectly elastic one);
(b) for a fixed of $\eta$, the energy loss is smaller for flattened clouds 
($L_y \gg L_c$).

We can rewrite  eqs. (9) and (23)
including the energy losses in the lateral outflow 
\begin{equation}
1-\epsilon={E_r \over E_k^i} \simeq (1-\eta_y)( 1-\eta ),
\end{equation}
which is valid for a head-on collision between identical clouds,
\begin{equation}
1-\epsilon={E_r \over E_k^i}\simeq f(\kappa,\beta)\left[{2-\xi \over
1+\kappa}(1-\eta_y)(1-\eta_2)+{\kappa\xi
\over 1+\kappa}(1-\eta_y)(1-\eta_1)\right],
\end{equation}
for a head-on collision between different clouds.
These results are obtained supposing that the lateral outflow 
is adiabatic.

\section{Numerical results}

In this Section we study cloud collisions via numerical simulations
adopting the same 1D model as in  the analytical approach.
The numerical code is based on a 
shock-capturing scheme (Yee, 1989) suitable to resolve the hypersonic
shocks that arise in the collisions. This  scheme is characterized by 
a nonlinear numerical dissipation term with an automatic feedback
mechanism which adjusts the amount of dissipation in any cell of the mesh
according to the shape of the actual solution. 
The spatial discretization adopted in our code is based on 
upwind-differencing and the solution is advanced in time 
using a Godunov-type method (solving a set of Riemann problems at any
cell interface forward in time).
To obtain a $2^{nd}$-order spatial accuracy we have used 
the TVD-MUSCL (Total Variation Diminishing-Monotone Upstream Scheme for
Conservation Laws) reconstruction technique.
We achieve $2^{nd}$-order accuracy in time  by using a two-step
Runge-Kutta explicit scheme.

%The code can be used with two different Riemann solvers: the approximate
%Roe Riemann solver and a more robust solver derived from the 
%Riemann solver by Osher (Osher \& Solomon, 1982; Abgrall et al., 1991).
%This solver is a 3-stage approximate solver in which the first stage
%computes the intermediate pressure and velocity assuming isentropic wave
%interaction. A second stage, based on the strong-shock relations,
%may be invoked to improve the first-stage estimate if the pressure jump
%across either wave is sufficiently large. The final stage is to
%select/interpolate the interface state from the set of left, right and
%intermediate states.
%
%The Roe Riemann solver, based on a linearization of the Riemann
%problem, may fail by predicting non-physical states with negative density
%or internal energy. This failure may happen because in flows where the
%dominant energy mode is kinetic the resulting internal energy may be
%negative. For this reason we have used the Roe Riemann solver only in
%the simulations with a non-zero magnetic pressure term  for which 
%the Osher solver is not applicable.
%As in \S~III, we will discuss separately the two cases
%of identical and different cloud collisions.

The simulation starts when the two clouds are in contact. The cloud and
intercloud density and temperature are computed from the corresponding phase diagram,
as described in Sec. 2. The external medium is initially put into motion,
with the same velocity as the clouds, to prevent the formation of a vacuum 
zone behind the cloud boundary due to the snowplow effect of the cloud motion.

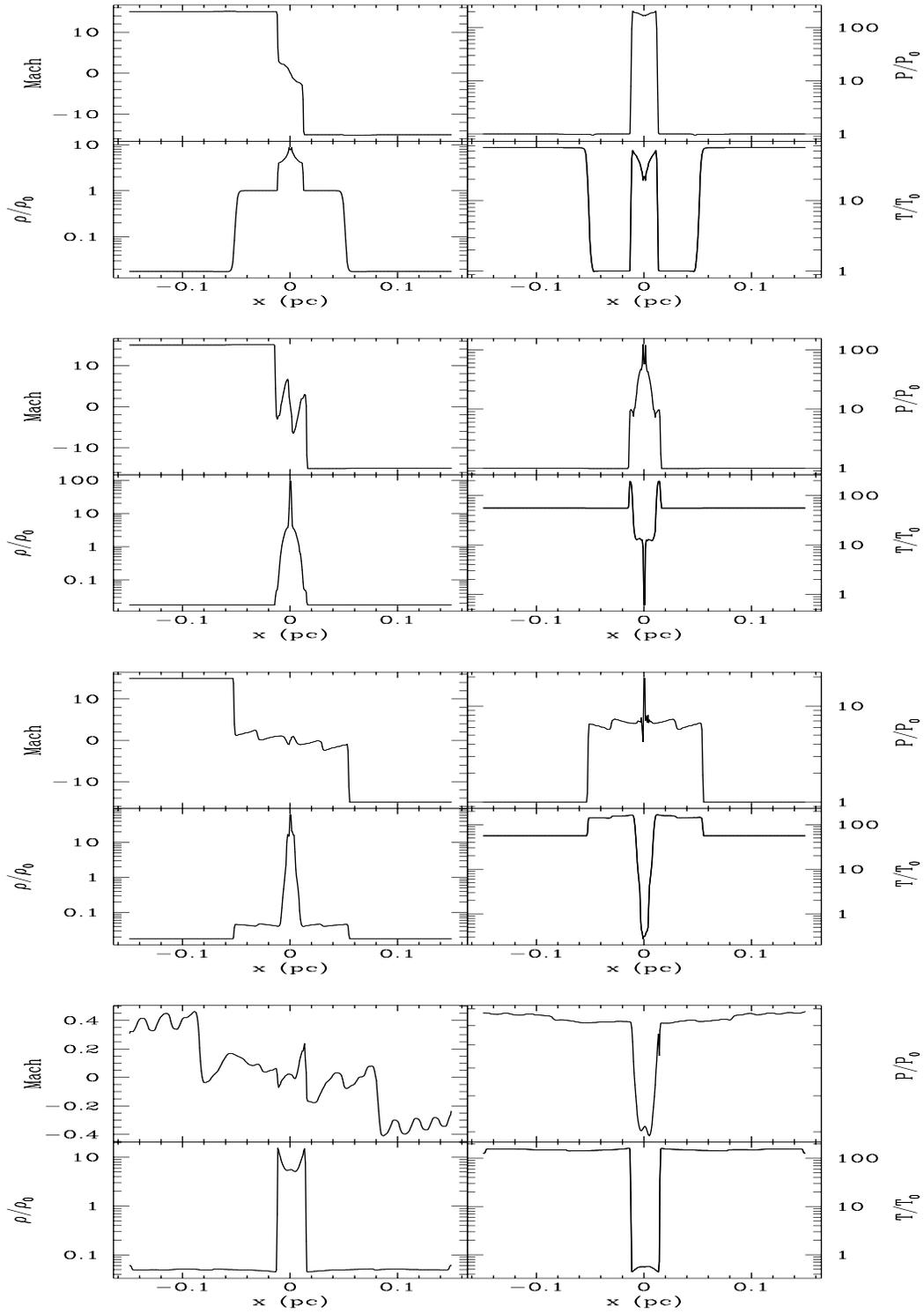
\begin{figure}
\caption{\label{fig3} Evolution of a collision between identical clouds
with relative velocity $v_r\simeq 19$ km s$^{-1}$, size $L_c=0.1$ pc,
$Z=D/G=1$. The panels show the density, velocity, pressure and
temperature at the evolutionary times $t\simeq$  0.5,0.98,1.6,4
$t_{coll}$, with $t_{coll}=L_c/v_c\simeq 10^4$ years.} 
\end{figure}

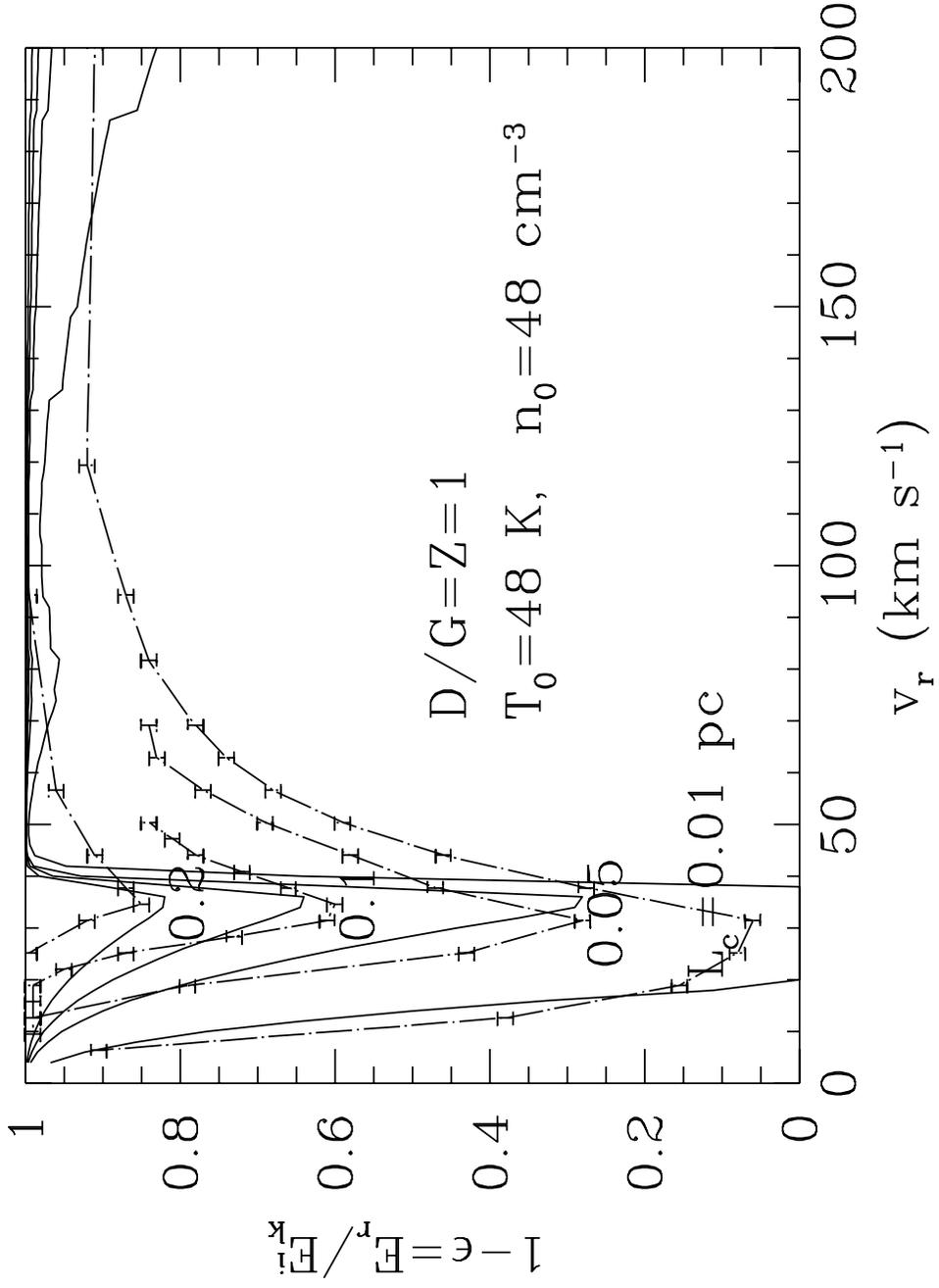
\begin{figure}
\caption{\label{fig4} Behavior of the quantity $(1-\epsilon)=E_r/E_k^i$, where
$\epsilon$ is the elasticity of the collision (see text) 
for a collision between two identical clouds of size $L_c$ shown by the labels, 
as a function of the cloud relative velocity
$v_r$. 
The gas
metallicity is $Z=1$. The solid curves represent the analytical expression for 
$1-\epsilon$ (where $\alpha=1/3$; see text); 
the points show the analogous results from the numerical simulations.
Vertical bars denote the estimated error in the simulation (see text).}
\end{figure}

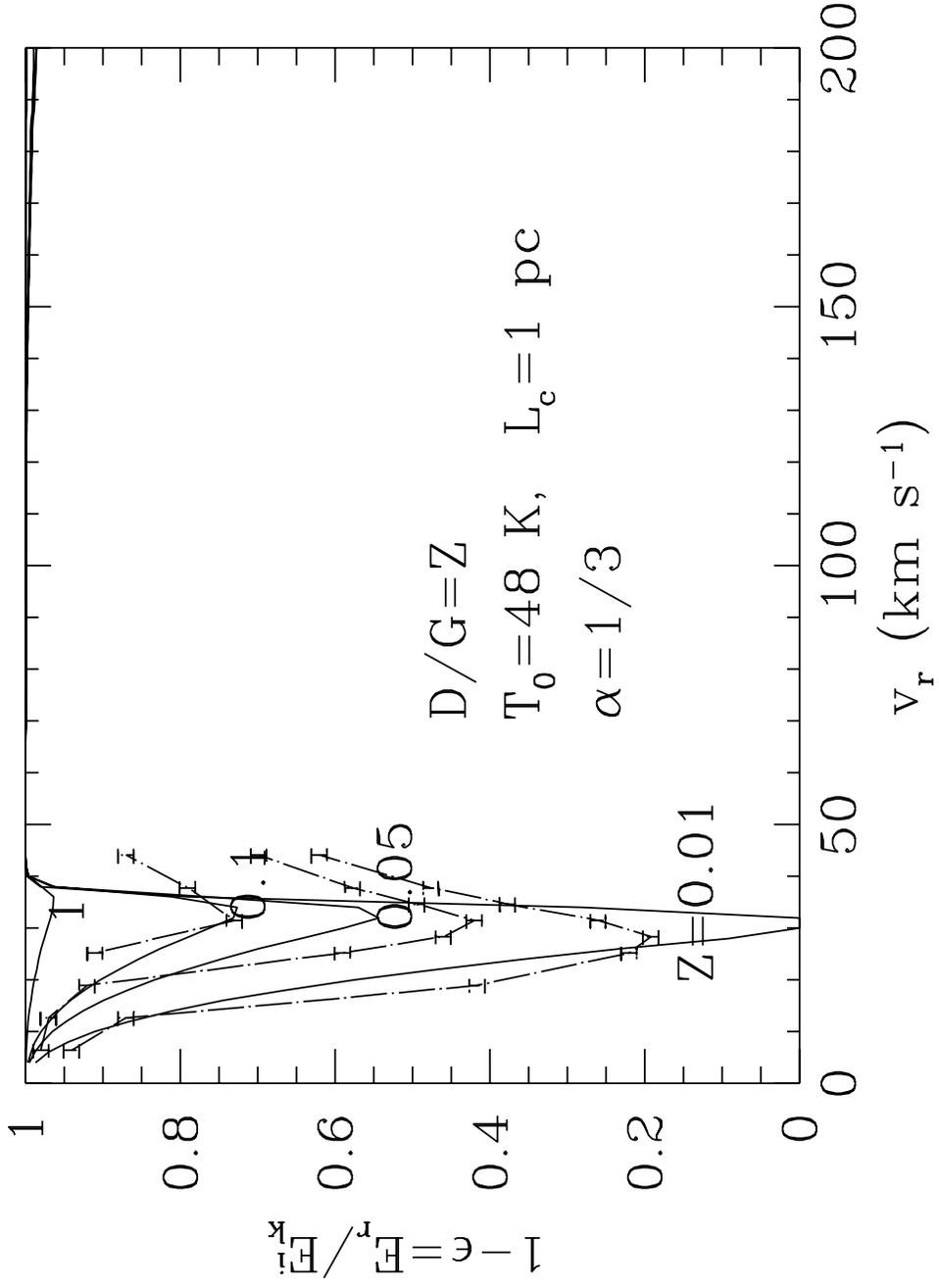
\begin{figure}
\caption{\label{fig5} Same as Fig. 4, but for
different values of $D/G$, and gas metallicity $Z$.
We assume $D/G =Z$, $L_c=1$~pc and $T_0=48$~K;
the cloud density $n_0$ is derived from the corresponding phase diagram.
Vertical bars denote the estimated error in the simulation (see text).}
\end{figure}

In Figure \ref{fig3} we show, as an example, the evolution of the collision 
of two identical clouds for the representative case  $v_r\simeq 19$ km s$^{-1}$, 
$L_c=0.1$ pc, $Z=D/G=1$. The three different phases discussed above 
are clearly seen in the various panels. At $t=0.5 t_{coll}$  
two shocks are already well formed and propagate through the cloud compressing
the gas. The pressure jump across the shock 
is $\propto v_s^2$ and the density jump is $\sim 4$
because the shock is still adiabatic. Then, the temperature of the
shocked gas starts to decrease due to radiative losses occurring in the
transition region (see Sec. 3).
At $t \simeq t_{coll}$ an expansion phase begins driven by
the hot gas near the cloud boundary whereas the gas in the cloud interior is
almost at rest. The shock propagates through the intercloud medium increasing both
its pressure and temperature. The cloud expansion is finally halted 
by the enhanced external pressure.

The behavior of the collision between two different clouds is initially
analogous to that of identical clouds (except that the CD 
is not at the rest in the LSR).
As the shock exits the smaller cloud, thus triggering its expansion phase,
the ram pressure of the larger cloud accelerates the CD and the velocity
of the shock in its interior decreases consequently.

Finally, we have performed some simulations of
collisions between identical magnetized clouds with the 
field lines parallel to the shock front. As expected, the main effect 
is a lower compression of the postshock gas, which enhances  the
value of $N_{cool}$. This has important consequences on the elasticity
of the collision as we will see in the next Section.

\section{Elasticity of the collisions}

In this Section we present (Figs. 3-7) the results concerning the 
elasticity of the collisions derived by analyzing the numerical simulations. 
The numerical results are compared with the appropriate analytical expressions
obtained in Sec. 3.
In order to achieve the best agreement between the two sets of results
we have fixed the value of the free parameter $\alpha$, introduced in
eq. (\ref{parametro_el}), to 1/3. 
With this choice we find a very good agreement also varying the cloud size 
and gas metallicity.
The error on the numerical calculation of $E_r$ is less than $1\%$ for all the runs,
and it is due to the fact that we integrate the radiative losses during the
simulation up to
the time at which the fractional variation of the integral is $< 1\%$.    
%The main result that we have derived is the simple analytical 
%relationship for the elasticity of the collision.
%For the centered collision of two identical clouds we have:
%\begin{equation}
%{E_r \over E_k^i}=(1-\eta),
%\end{equation}
%where the ``elasticity parameter'' $\eta$ is:
%\begin{equation}
%\eta \propto {N_{cool} \over n_0 L_c}~~~ {\rm with}~~~ 0 \le \eta \le 1,
%\label{concl1}
%\end{equation}
%Since $N_{cool}=n_0 v_s t_{cool}$, with $v_s \propto v_r$ shock
%velocity, $t_{cool} \propto v_s^2/[n_0 \Lambda(v_s,Z)]$ typical time of
%cooling of the shocked gas and $\Lambda $(ergs cm$^3$ s$^{-1}$) gas
%emissivity, we have:
%\begin{equation}
%\eta \propto {v_r^3 \over n_0 L_c \Lambda(v_r,Z)},
%\end{equation}
%that, for one-dimensional flows, is the collision elasticity $E_k^f/E_k^i$,
%where $E_k^f$ is the kinetic energy of the clouds after the collision.
%The results of the numerical simulations, performed with an
%one-dimensional TVD MUSCL shock capturing code, agree rather well with the
%analytical expressions if we fix the proportional parameter of the eq.
%(\ref{concl1}) to 1/3.
 
In the following we will discuss the detailed dependence of $E_r/E_k^i$
on the various collision parameters, namely $v_r, L_c, Z, \kappa, \beta, B$.
\\
$\bullet E_r/E_k^i(v_r)$ (Fig. 3) 
If the relative velocity of the clouds, is $v_r\ll v_m$ or $v_r \gg
v_m$, with $v_m \simeq 30$ km s$^{-1}$, the collision is
approximately inelastic (\ie all the initial kinetic energy of the
clouds is dissipated radiatively). In a collision between identical
clouds, this occurrence does not depend on any other parameter of
the collision.
For $v_r$ in the interval ($v_m-\delta v,v_m+\delta v$)
where $\delta v \sim 10$ km s$^{-1}$, the energy dissipation in the 
collision decreases and for $v \simeq v_m$
it reaches a minimum (\ie the elasticity is maximum).
The value of $v_m$ is slightly dependent on the metallicity 
$Z$ of the gas.
This behavior of the elasticity as
a function of the relative velocity can be understood as follows.
We recall that 
$N_{cool}=n_0 v_s t_{cool}$, with $v_s \propto v_r$;
since $t_{cool} \propto v_s^2/[n_0 \Lambda(v_s,Z)]$, 
then $N_{cool}$ is the product of two terms: the first,
$v_s^3$, increasing, the second, $\Lambda^{-1}(v_s,Z)$, decreasing with $v_s$. 
Thus $N_{cool}$ has a maximum and the same is true for $\epsilon \propto N_{cool}$. 
Differentiating $epsilon$ with respect to $v_r$ we find that $\epsilon$ 
is maximum when:
\begin{equation}
{\partial \Lambda (v_r) \over \partial v_r}=3,~~~~~~{\rm or}~~~~~~
{\partial \Lambda (T) \over \partial T}={3 \over 2}, 
\end{equation}
\\
$\bullet E_r/E_k^i(L_c)$
The elasticity of the collision increases when the cloud size
is decreased because $\eta \propto L_c^{-1}$ (Fig. 3).
\\
$\bullet E_r/E_k^i(Z)$
The elasticity of the collision increases when the gas metallicity 
is decreased because $\eta \propto \Lambda^{-1}$ and 
$\Lambda$ is directly proportional to $Z$, as shown by Fig. 4.
Interesting enough, collisions between clouds for which the 
relationship $ZL_c^2 \approx {\rm const.}$ holds, are equally elastic, 
for a given value of $v_r$. The physical explanation can be found 
from an inspection of eq. (\ref{eq_eta1}) and recalling that, to a first
approximation, $\Lambda \propto Z$ and the equilibrium density $n_0$ 
of the cloud is $\propto Z^{-1/2}$. This relationship is illustrated by Fig. 5.
\\
$\bullet E_r/E_k^i(\kappa,\beta)$
When the cloud size ratio $\kappa=L_1/L_2$ and the cloud velocity ratio
$\beta=v_1/v_2$ are not equal to one (\ie different clouds), the maximum
kinetic energy that can be dissipated in the collision is equal to the
kinetic energy relative to the center of mass reference frame.  Hence, the
collision between two different clouds will be, in general,  less
dissipative with respect to the identical cloud case. 
Moreover, while the shock velocity in the smaller cloud is always 
$\propto v_r$, the shock will be slowed down in the larger cloud
for the reasons discussed in Sec. 4. 
This will produce an additional minimum of $E_r/E_k^i$ at $v > v_m$ as
shown by  Fig.\ref{fig6}. The analytical approximation in that region
tends to overestimate the elasticity by about 10%.
\\
$\bullet E_r/E_k^i(B)$
The effect of an uniform magnetic field $B_0$ in the cloud, is to
decrease the energy dissipation (\ie increase the elasticity) of the
collision. Since the magnetic pressure limits the compression of the
postshock gas; as a consequence,  $N_{cool}$ increases along with the elasticity
of the collision (see Fig.\ref{fig7}).
In Figure~\ref{fig7} we show the behavior of $(1-\epsilon)$ vs. 
${\cal B}=P_B(0)/P(0)$ (the ratio of magnetic and thermal pressure at $t=0$).

\section{Summary and discussion}

We have presented a first step towards the study of the collisions between 
diffuse interstellar clouds in a multiphase
medium, with particular focus to the dissipation of the kinetic energy of
the clouds (described through the elasticity parameter, \ie the ratio of the
final to the initial kinetic energy of the clouds)
as a function of the parameters of the problem such as the cloud relative
velocity, gas metallicity, cloud mass and velocity ratio, and magnetic field strength.

The problem has been studied both analytically, obtaining approximate, albeit
handy, relations valid for a wide range of  parameter variation, and numerically,
by means of a 1D {\it shock capturing} TVD numerical code. 
The comparison between the two approaches has been explored in detail (see
Figs. 3-7); we conclude that the agreement is very good. 

The following points summarize our main results:
\\
$\bullet$
The kinetic energy dissipation in cloud collisions is minimum 
(\ie the collision elasticity is maximum) for a cloud relative velocity
$v_r \simeq 30$
km s$^{-1}$.
\\
$\bullet$
The above minimum value is proportional $Z
L_c^2$ (where $Z$ is the gas metallicity and $L_c$ is the cloud
size): the larger is $ZL_c^2$ the
more dissipative (\ie more inelastic) the collision will be.
\\
$\bullet$
We find that the energy dissipation decreases (\ie elasticity
increases) when the magnetic field
strength, and mass ratio of the clouds are increased and the metallicity is
decreased, respectively. 

The previous results have been obtained assuming:
(i) centered collisions: we have investigated a simple 1D model of the collision only;
the extension to the analogous 2D problem will be presented in a future paper.
However, we have already explored the effects of the lateral outflow analytically.
(ii) ionization equilibrium:
this holds only approximately if the temperature of the post-shock 
gas if $T< 3\times 10^4$~K and $Z> 0.01$
(iii) the ionization of the pre-shock gas by the 
radiative precursor (Shull \& McKee 1979) which could modify the shock structure 
is neglected.  (iv) clouds are not self-gravitating. 

Our results, mainly aimed at deriving the dependence of the elasticity
on a wide range of parameters characterizing the initial conditions of
the collision in a multiphase medium, do not allow us to draw any firm conclusion
on the final fate of the remnant of the collision (shattering or coalescence). 
In order to make progresses a 2D study is required. Nevertheless, our 
perception from the 1D study presented here is that the presence of a  
magnetic field must be invoked in order to prevent the shattering of the clouds
found by Klein \etal (1995).
A related point is the possible 
phase transition from the CNM to the WNM associated with the collisions,
which could result in a net mass and energy exchange from the cold to the warm
phase. This could be relevant as far as the overall thermal equilibrium of the ISM
is concerned.
 
Cloud collisions might be responsible for the buildup of the observed mass
spectrum ($N(m) \propto m^{-2.14}$, Dickey \& Garwood 1989) of diffuse clouds and for
the formation of molecular complexes. Das \& Jog (1996) study the evolution
of molecular clouds under the effect of collisions and local gravitational 
interactions. Their recipe for the collision is a simple extrapolation of the
results by Hausman (1981), \ie coalescence occurs in a subsonic collision,
fragmentation is the product of a supersonic one. However, most of the standard 
buildup models of diffuse clouds -- inspired by the pioneering work of Field 
\& Saslaw (1965) --
assume that collisions are inelastic, a hypothesis that is too simplistic by far,
as already made clear by our 1D results. Also, as Jungwiert \& Palous (1996) 
have pointed out, the elasticity of cloud collisions is a key parameter for the
process of formation of multiple rings in disk galaxies. 

In a cosmological context, collisions between primeval clouds could be important
for two reasons. On the one hand, the nonlinear evolution of primordial 
fluctuations is thought to generate a very clumpy state of the intergalactic
medium. One can expect these protogalactic seeds to move through the
background gas, collide and eventually coalesce; during this process
part of their orbital momentum can be transformed into spin momentum of 
the merger (Chernin 1993).
On the other hand, energy dissipation by means of cloud collisions in a 
forming galaxy is found to produce flattened systems (Gott \& Thuan 1976, Larson
1976, Abadi \etal 1990) and the disks of spiral galaxies.
It is thus important to fully understand how the energy dissipation is affected
both by the gas metallicity, which is supposed to be rather low, 
and by the phase structure of the ISM of these primeval objects. 
Our calculations provide a necessary ingredient for this type of studies. 

\vskip 2truecm
We are grateful to S. Balbus, J. Dickey, G. Field, R. Klein, C. McKee, and M. Shull
for stimulating discussions. 

%.......................................................................

\newpage

%\figcaption[fig1.ps]{The extinction curve resulting from the adopted
%dust model. Also shown are the contributions of the single components.
%The color excesses for the single materials are divided by the value of
%E(B-V) for the total mixture. \label{estinzio}}

\begin{table}[h]
\begin{center}
\begin{tabular}{|l|l|c|}
\hline
              &     & \\
     Dominant cooling processes & Notes & Ref. \\
              &     & \\
\hline
        Cooling by fine-structure lines:      &  &    \\
        CII(158$\mu$m)      & Impacts with H$^0$ and $e^-$& 1  \\
        OI(63;44$\mu$m)    & Impacts with H$^0$,$e^-$ and H$^+$ & 2  \\
        FeII(26;15$\mu$m),SiII(34.8$\mu$m)  & Impacts with H$^0$ and $e^-$&
	3\\
        Cooling by metastable lines:      &  Impacts with $e^-$  & 3 \\
        CII(2326$\AA$)      &    &  \\
        OI(6300;6363$\AA$),OII(3229;3726$\AA$)      & OI most important
	coolant &   \\
        NI(1.04$\mu$m),NII(6548;6583$\AA$)      &    &  \\
        FeII(5.34;4.12;1.26$\mu$m)      &    &  \\
        SII(6731;6717$\AA$)      &    &  \\
        SiII(2240$\AA$)      &    &  \\
        Cooling by highly ionized elements:  &  Impacts with $e^-$ & 4,5  \\
        CIII,CIV,OIII,OIV,OV      &  Important at  $T>10^5$ K& \\
\hline
\end{tabular}
\end{center}
\caption{\label{processi_cool} Dominant cooling processes by metal
lines. Reference:
(1)Wolfire {\it et al.}, 1995;(2)P\'equignot, 1990;(3)Hollenbach \&
McKee, 1989;(4)Gaetz \& Salpeter, 1983;(5)Shull \& van Steenberg,
1982.}
\end{table}
\vfill\eject

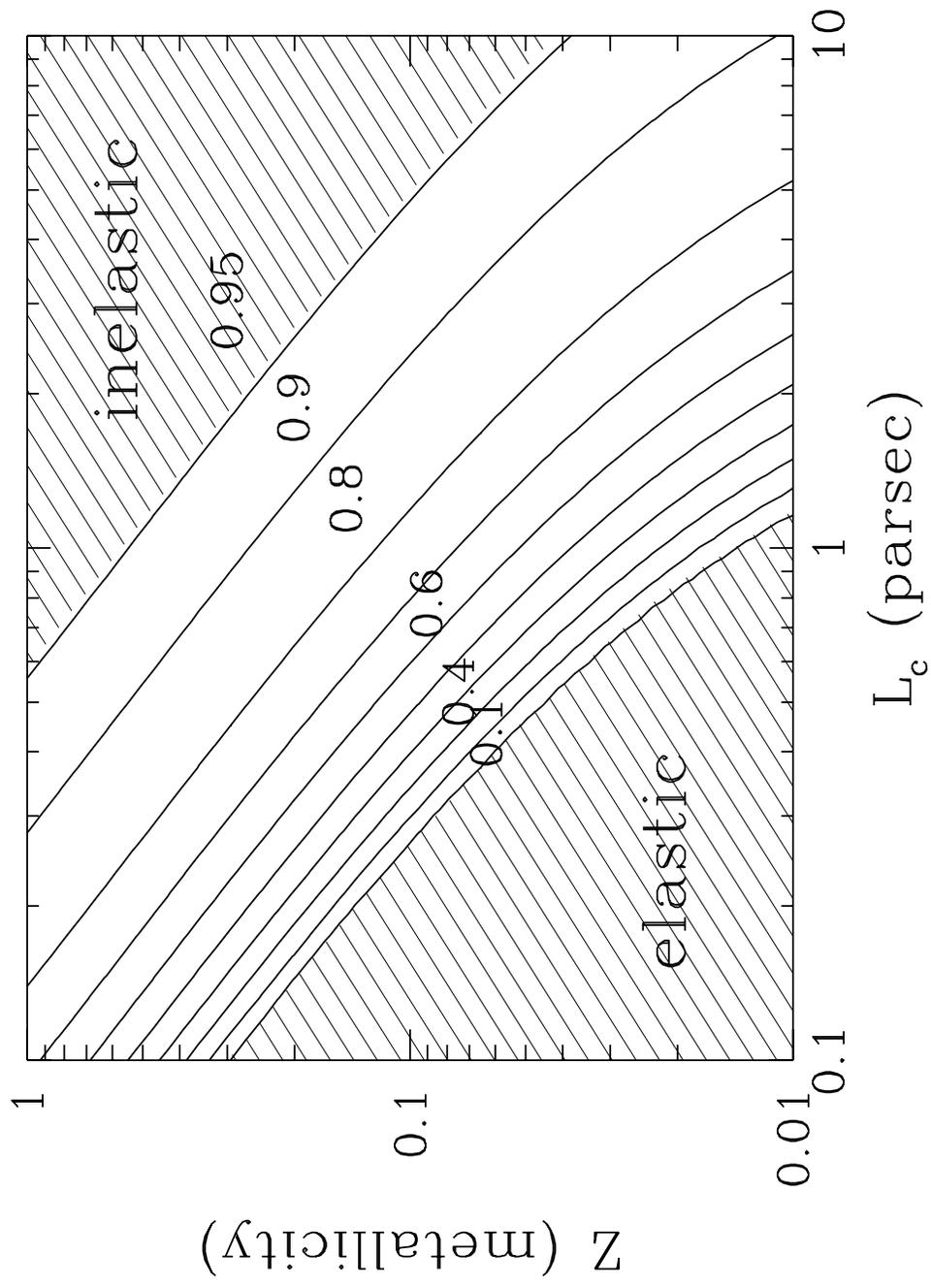
\begin{figure}
\caption{\label{conc_f2} Contour levels of the function
$1-\epsilon=E_r/E_k^i$ in the plane  $L_c-Z$ for a  relative
velocity  $v_r=30$ km s$^{-1}$.}
\end{figure}

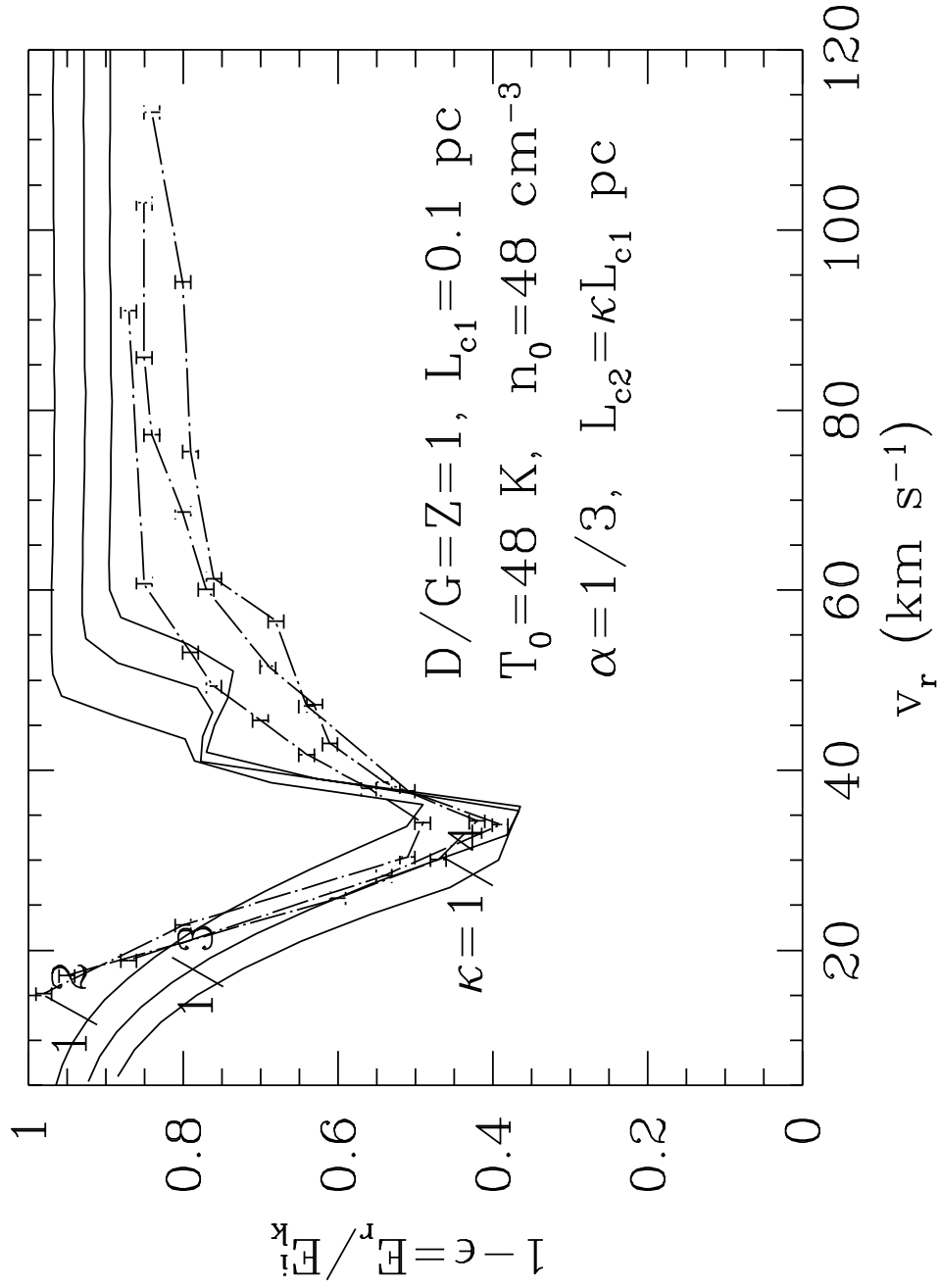
\begin{figure}
\caption{\label{fig6} Same as Fig. 4, but for a  
collision between different clouds as a function of the cloud relative
velocity
$v_r$; the curves refer to different values of the cloud size
$\kappa=L_{c1}/L_{c2}$ and velocity  $\beta=\vert v_1/v_2 \vert$ ratios; 
we have assumed $\beta^2=1/\kappa$ so that the clouds have the same
initial kinetic energy.
The points show the analogous results from the numerical simulations.
Vertical bars denote the estimated error in the simulation (see text).}
\end{figure}

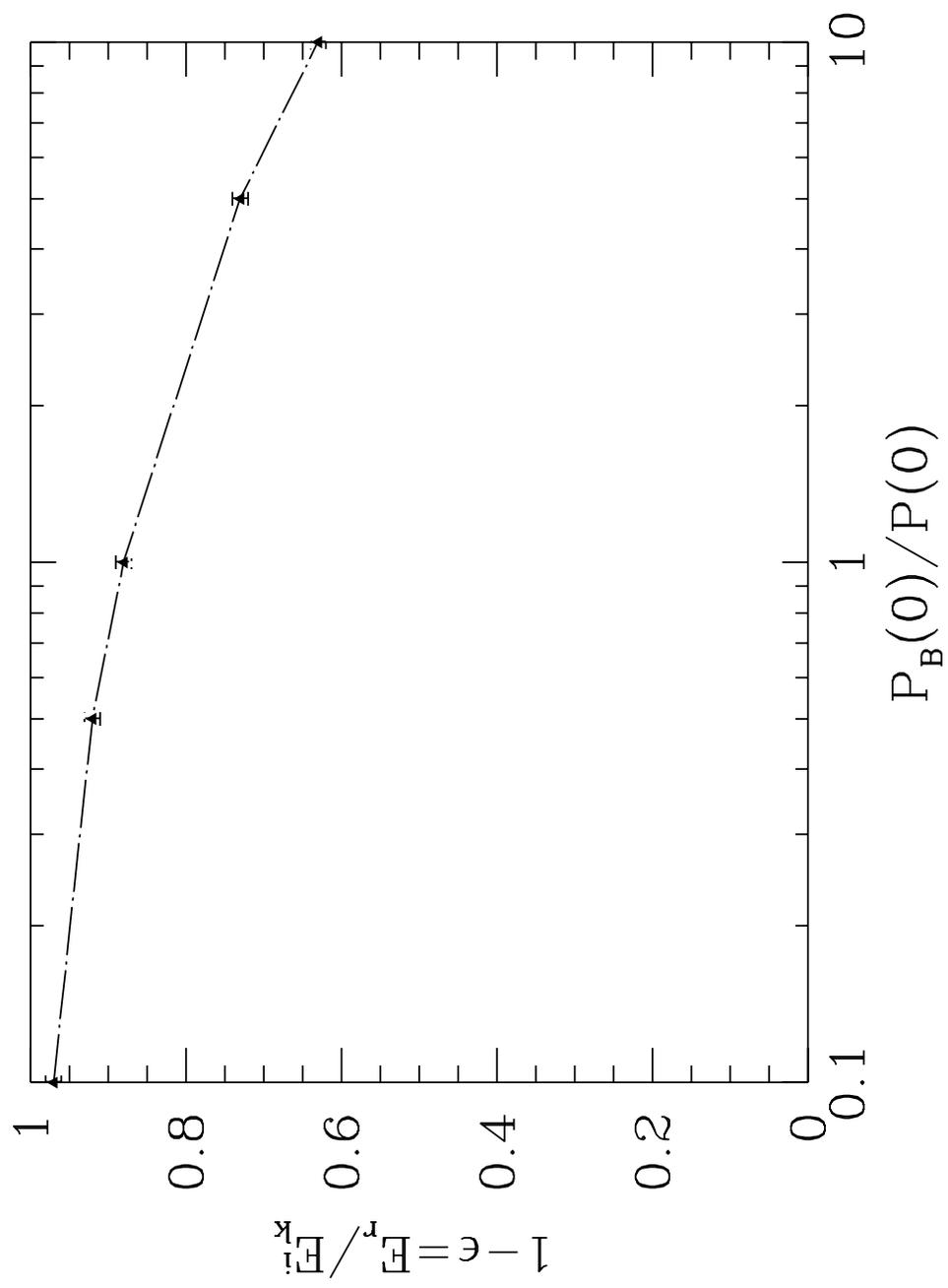
\begin{figure}
\caption{\label{fig7} $1-\epsilon=E_r/E_k^i$ curves plotted as a function of
the initial ($t=0$)  ratio between the magnetic and thermal pressure $P_B(0)/P(0)$
in the cloud.
The points shown in the picture are computed for: $P_B(0)/P(0)=0.1;0.5;1;5;10$
(\ie $B_0\simeq 1;2;3;6;9\mu$G). The collision parameters are the same as in  
Fig.~\ref{fig3}.
Vertical bars denote the estimated error in the simulation (see text).}
\end{figure}

\end{document}